# Enhancement of the catalytic activity upon the surface of strongly disordered hollow Pt nanoparticles: a First Principles investigation.

Olivier Le Bacq[1]

Univ. Grenoble Alpes, CNRS, Grenoble INP, SIMaP, F-38000 Grenoble, France

date: 11/24/23

*A series of First Principles calculations is undertaken to characterize and explain the enhancement of the catalytic activity of oxygen on top of very disordered nanomaterials of Pt. As the adsorption of OH fragment on top of the surfaces is known as the limiting factor in the Oxygen Reduction Reactions (ORR) process in these systems, our calculations propose to determine the influence of the local geometry of the various sites on the adsorption energy of OH in order to discover a simple descriptor allowing to predict the reactivity at these surfaces as a function of their morphology and strain. For this purpose, the geometry of Pt slabs with various thickness (3, 5 and 7 atomic layers) including a large number of point defects are optimized in order to generate a very rich catalog of inequivalent sites of reactivity on both surfaces of the slabs. Given the very large distortion of the geometry of the sites, these latter had to be categorized into several classes for which the behavior with respect to catalytic activity is determined. A new descriptor taking into account the distortion of the geometry of the sites is introduced, allowing to recover the linear dependence of the adsorption energy of OH with respect to the effective coordination number of the sites, as observed in highly symmetric and planar surfaces of Pt.*

---

1 Corresponding author: olivier.lebacq@grenoble-inp.fr

**Introduction.**

Electrochemical storage and conversion systems such as electrolysers and fuel cells will be major actors. They can store electrical energy into chemical energy via water electrolysis (for example into the H-H bond of the $H_2$ molecule), and convert back this chemical energy into electrical energy via a fuel cell when needed[1]. A major challenge in these systems is to find highly active and stable materials to sustainably catalyse the electrochemical reactions[2].

The research in electrocatalysis has focused on binding electrochemical reactions intermediates in a balanced way (that is neither too strong nor too weak). A prominent example is the oxygen reduction reaction (ORR), the cathodic reaction in low temperature proton-exchange membrane fuel cells (PEMFCs). Weakening the chemisorption energies of the ORR intermediates ($OH_{ads}$, $OOH_{ads}$ and $O_{ads}$), and thus enhancing the overall kinetics proved experimentally possible on platinum (Pt, the most active pure catalyst for the ORR in acidic media at medium temperature $T \sim$ 353 K) surfaces alloyed with an early[3] or late [4] transition metal (referred to as M in what follows). The findings made on extended surfaces were then transposed on nanometre-sized particles, and led to the emergence of different catalyst's architectures including pure PtM alloys [2.a-5], Pt skin/skeleton PtM-type catalysts[6], Pt-monolayer-type catalysts,[7] dealloyed PtM nanoparticles,[8] preferentially-shaped nanoparticles,[5.b-9] nanostructured thin films[10], and porous PtM catalysts[11] .

In this context, we pursue our investigation of distorted Pt surfaces as an elementary model for the walls of the Pt nanomaterials. Following a detailed analysis of the geometry of these systems that can be found in Ref. [12] and an outlining of the role played by the surface distortion (Ref. [13]) as a unifying descriptor, the present work tries to determine an implicit descriptor taking into account the local geometry of the catalytic sites that could allow to rationalize the behaviour of ORR process on the distorted surface of the nanomaterials. Following the works of Calle-Vallejo et al. (Ref. [14]), we consider the adsorption of the OH fragment, occuring in the electrochemical reactions intermediates of the ORR, as the limiting factor, mainly piloting the efficiency of the catalytic activity: a too binding OH fragment located on a specific surface site will poorly contributes to the ORR since its inability to leave that site. To the opposite, a local geometry presenting an unfavourable situation for OH adsorption will hardly allow OH to adsorbate on the surface. Here, we propose to calculate the adsorption energies of OH for a large variety of local distorted geometries in order to find a descriptor allowing to classify the sites with respect to the catalytic activity.

After relaxation of the slabs, DFT calculations permit to access the full distribution of distances around every surface sites. This work aims to use this knowledge to form a geometry-dependent descriptor, capable of predicting the adsorption energy of OH without having to consider the impossible DFT calculation of all the adsorption energies on every sites. For this purpose, we realized that the sites could be recast into families or types that shall be describe in details in paragraph 2.a. For selected sites of these families, a series of DFT calculations of the adsorption energy of OH were operated and stored in a data base. This allowed an acceptable and feasible amount of calculations permitting to fit the behaviour of this energy with respect to a suitable variable, representing the aspect of the local geometry of the site on which the adsorption is considered. We establish this law of behaviour in paragraph 4 before illustrating its scope.

1. **Methods.**

   a. Strategy of calculation.

Although made of an assembly of fcc nanocrystals of Pt, orientated both [111] and [100] in order to close the nanoparticles, our first investigation of their wall will focus on the most stable [111] surface (hexagonal symmetry). Slabs of 3, 5 and 7  8x8 layers involving 200 to 448 atoms were prepared (see Fig. 1). The modeling of the step-by-step elaboration of the nanoparticles being out of reach by means of our DFT calculations, we have mimicked the effects of such an elaboration by introducing an inhomogeneous density of points defects in the wall as suggested by the experimental mechanisms of elimination of particles of substrate (Ni) during the gradual making of the nanomaterial. The vacancies are introduced at random in the slab with the constraint to be found in larger amount in both first layers defining the surface than deep inside the wall (one layer for the 3 and 5 layers slab and 3 layers for the 7 layers slab since two surfaces have to be considered). This latter point, suggested by the experiments, conveys that atoms of substrates are more easily evacuated at the surface, leaving points defects in this later during the process, than inside the wall under self-construction, in which Ni atoms are extracted too, but in a lower extent, while a fraction of them remain trapped (about 12%). In the present study, the Ni atoms are considered as fully evacuated and will not be taken into consideration.

As the amount of vacancies involved in the process of elaboration of our nanoparticles is unknown, the number of point defects we introduced in the slabs were considered as a variable in our problem : for this purpose, in the case of the 5 layers slab which is considered as our reference, , the density of vacancies, denoted $\rho_v$ in the followings, were varied from a low value of 30 % to the upper limit of 50 %, by successive increase of 5 %. Below 30 %, the phenomena of reconstruction of surface were found similar to the one encountered in the 30 % case and 50 % found large enough to explore all the variety of morphologies made possible by the massive introduction of vacancies in these systems.

Such a procedure led after relaxation to the determination of the equilibrium structure (surface and inner structure) of the walls of the Pt nanoparticles for various thickness, obtained under varying conditions: the larger $\rho_v$, the more our model of wall is supposed to represent an actual wall resulting from a large and penetrating diffusion of atoms of the Ni substrate during its elaboration. Low densities of defects could result from a slow and moderate diffusion of Ni inside the Pt multilayers under forming.

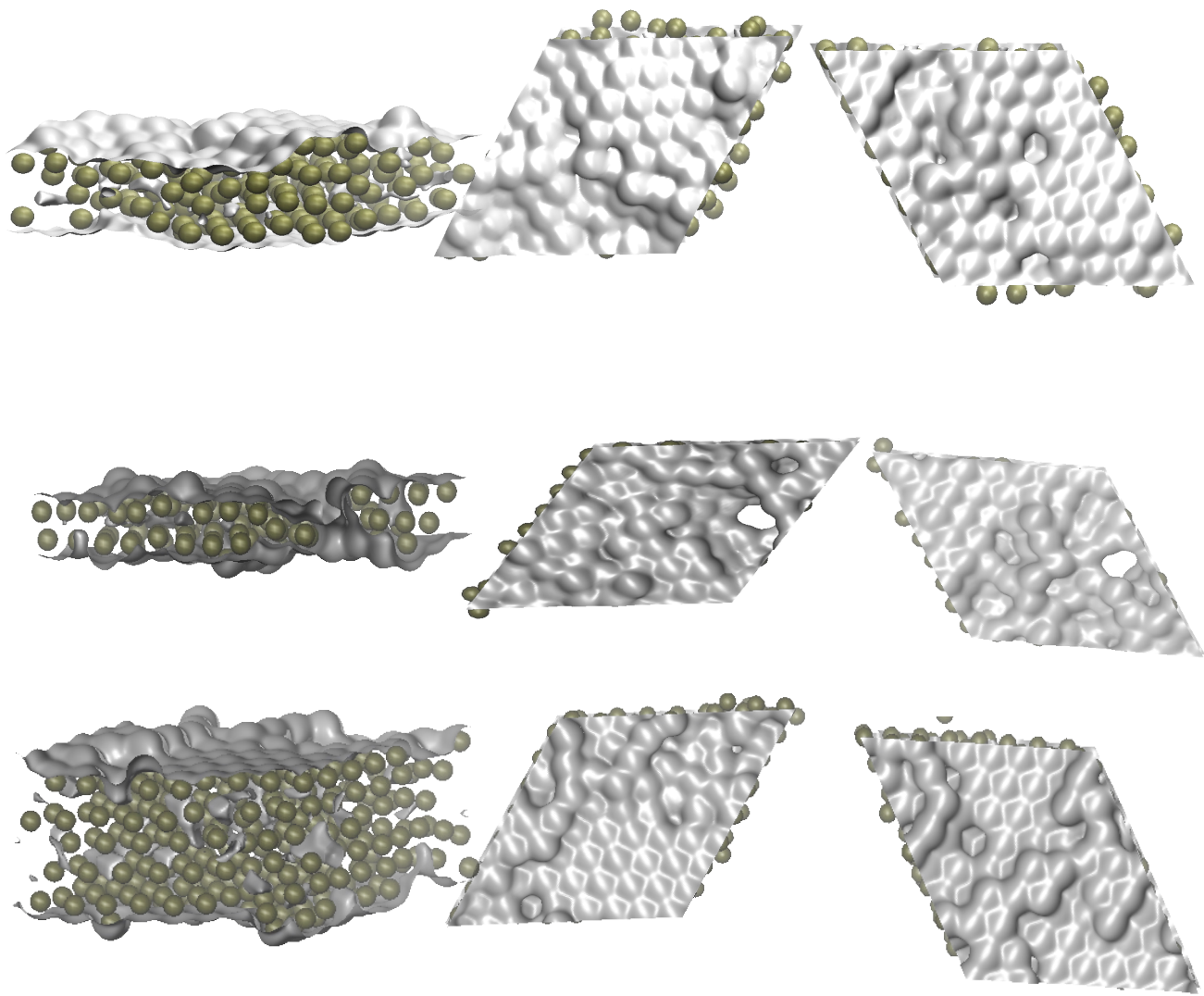

*Figure 1: Typical geometries obtained after relaxation of our 5, 3 and 7 layers slabs. For each line, a perspective view as well as the view of the top and bottom surface of the slab is shown.*

b. Settings of the calculation.

To compute structural and electronic properties of our slabs, we applied the projector augmented wave (PAW) method [14] as implemented in the Vienna ab initio simulation package (VASP) program [15]. The method, based on the density-functional theory (DFT) is an all-electron method allowing a correct description of the valence wave functions and its nodal behavior without any shape approximation on the crystal potential. Calculations were carried out using the general gradient approximation (GGA) with the Perdew–Burke–Ernzerhof (PBE) functional [16].

Numerical integrations in the Brillouin zone were performed by means of the Hermite-Gaussian method. A single k-point mesh (Γ point) was found enough for total energies of the investigated compounds to be converged within 10-3 eV. To optimize the geometry of cells, we have performed a volume optimization of the computational cell as well as internal relaxations of the atomic positions using the Hellmann-Feynman theorem. Since the large amount of defects we introduce at random in the slab is expected to break the overall symmetry, this later were switched off in order to allow the forces on the ions as well as stress tensor to be calculated without the constraint to follow the space group operators imposed by the initial geometry. We have considered the atomic coordinates are fully relaxed as the size of the individual forces were less than 0.001 eV/Å. We applied a residual minimization method for this purpose.

A series of brief first principle molecular dynamics simulations aiming at heating the so-obtained equilibrium geometries until 700 K and immediately followed by a quench (relaxation at 0K of the heated conformation) were performed in order to check the stability of our structures.

2. **A specific study case: 5 layers of Pt orientated [111] with $\rho_v$=40% .**

This paragraph is devoted to the detailed study of the slab made of 5 layers of Pt and $\rho_v$=40%. This particular slab, selected among all the ones we have investigated will allow us to extract and comment the major features of our Pt nanoparticles. It has to be viewed as a reference and starting point from which further studies can be carried out. Indeed, this slab were selected because it presents the advantage to display a rich variety of surface geometries. Moreover, the acceptable number of atoms it contains (380 atoms) allows to perform the large number of very time-consuming calculations needed to get a first valuable insight into the physics and chemistry of our compounds. Our results will be generalized and augmented in section 3 by making use of the data of our whole set of slabs of lower (3 layers) and upper (7 layers) dimensions.

a. Categorization of the sites.

Despite all the sites of surface are found non-equivalent due to the strongly disordered character of the surface and inner structure of our compounds, a categorization of these sites into groups of similar behavior has to be attempted in order to get a clear picture of their space distribution into the surfaces (are the sites distributed at random or in domains) as well as their response towards the ORR activity. To achieve our goals, a series of variables defining such a categorization has to be chosen: these variables have to permit to recast the sites into classes of atomic sites of similar structural parameters, and, here, hope is at this stage, that each class will display a similar behavior regarding ORR activity. The analysis of the local geometry of numerous sites of surface belonging to slabs of Pt of various thickness and densities of defects (see Figure 2-h/ for a typical example and figures of Ref. [12] for additional examples) makes appear that, whatever the coordination number of the site maybe, some sites display a character of moderate symmetry (the first neighbors interatomic distances are tightly gathered around their average value, $\lambda$, and some others present a wide distribution of interatomic distances around the average value (very disordered situation). From these observations, we found it legitimate to categorize the sites for each CN according to their average first neighbor distance, $\lambda$, and the width of the distribution of interatomic distances, $\delta$, defined by the difference of the shortest distance with the longest one, enclosed in a sphere of radius 2.9 Å around the considered site (Figure 2-h/ displays a graphical representation of these variables). However, local geometries are so various that a mere plot of the points representing the sites in the ($\lambda$,$\delta$) map is found insufficient to extract efficient classes of sites displaying similar structural parameters; and we had to introduce acceptance limits on both $\lambda$ and $\delta$ to neatly separate distinct classes of behavior among the sites. These latter are incorporated by replacing the point of representation of a given site in the (λ,δ) map by a normalized Gaussian function of both λ and δ variables for which full width at half-maximum coincides with the acceptance limits. This way, a distribution function is postulated in the (λ,δ) map. i.e.

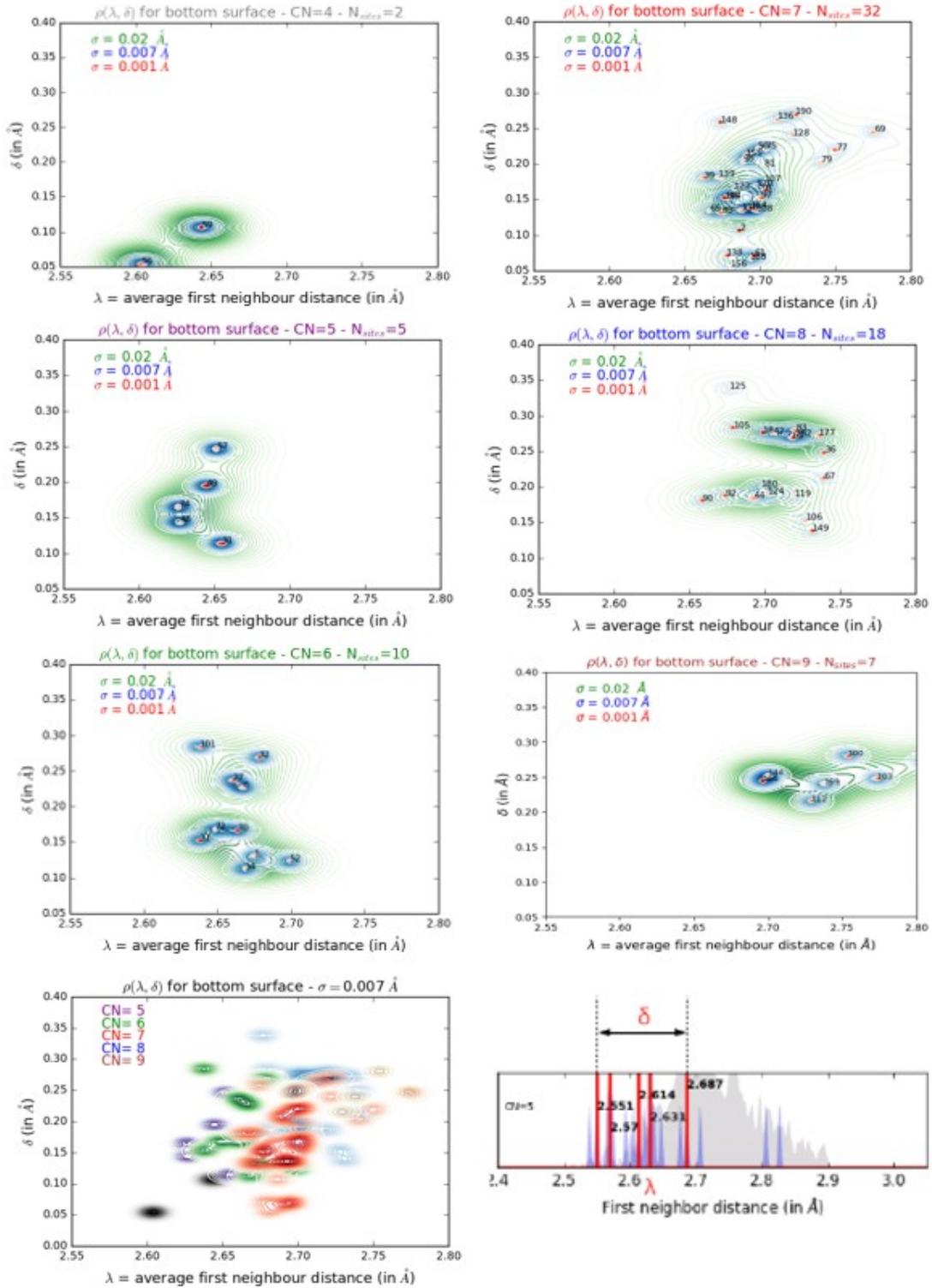

*Figure 2: Distribution of surface sites in the (λ,δ) map resented for each CN for the bottom surface of the 5 layers slab with ρ$_v$=40%. A final plot shows a superposition of them allowing to compare their relative location in the (λ,δ) map. Besides, we provide a graphical definition for λ and δ (see text for further explanation).*

$$\rho(\lambda,\delta)=\sum_{i=1}^{N}\frac{1}{N 2\pi \sigma_\lambda \sigma_\delta} e^{\frac{-(\lambda-\lambda_i)^2}{2\sigma_\lambda^2}} e^{\frac{-(\delta-\delta_i)^2}{2\sigma_\delta^2}} \qquad (1)$$

where $\sigma_\lambda$ and $\sigma_\delta$ are the acceptance limits corresponding to the discrete values of the average first neighbor distances, $\lambda_i$, and spectral width, $\delta_i$ of site i as obtained by our DFT calculations, respectively. N is the number of sites in the surface. Figure 2 shows the function ρ(λ,δ) for three values of $\sigma_\lambda = \sigma_\delta$ = 0.001, 0.007 and 0.02 Å. The lowest value (red dots in Figure 2) separates too much the sites and does not allow to define any category. With an increasing σ, domains start to develop (blue area) and for σ=0.02 Å, well defined classes of site can be observed (green area). These plots of density in the (λ,δ) map reveals that:

i/ Most of the surface sites display a CN of 6, 7 and 8. The sites with remaining CN (3,4,5 and 9) can be considered as negligible with respect to the previous ones. As we shall see in the followings, these latter account for the very extreme local geometries (sites located deep inside holes and at the top of *hills*).

ii/ for a fixed CN, the average first neighbor distance hardly varies from one site to another.

iii/ two types of population are obtained for each sites of same CN: one with small values of δ (rather ordered sites) and one with larger values of δ (rather disordered sites). We have fixed at $\delta_c$=0.2 Å, the critical value of δ separating both these distinct populations. This choice may appear quite arbitrary but was actually motivated by the very clear frontier obtained graphically in Figure 2 for CN=6 and 8, also observed on density plots of the other slabs for the same and others CN (not represented).

iv/ the average first neighbor distance increases together with the CN of the site, as revealed in the plot of Figure 2 displaying the superimposition of all the densities in the (λ,δ) map: despite an unavoidable overlapping, distinct colored strips are clearly noticeable. This result was actually expected: the more numerous the neighbors are, smaller their input of charge must be to close the electronic shell of the central site: they have to move away to achieve it, so that their average neighbor distance increases.

Thus, and from now on, the various sites may be considered as belonging to distinct classes of atoms and the following color code has to be kept in mind to distinguish in between them in the continuation of this paper: sites with CN=5 will be plotted in violet, CN=6 in green, CN=7 in red and CN=8 in blue. The atoms circled in black will reveal the more ordered character of the site ($\delta_i < \delta_c$), the most disordered sites will not be circled: thus, a sub-categorization is operated for each value of CN. In the rest of this paper, we shall denote sites with $\delta_i < \delta_c$ as S-site (fcc-symmetry) and sites with $δ_i < δ_c$ as D-sites (Disordered geometry). Fig. completes to illustrate these notations.

   b. A first insight into the dispatching of these categories within a surface.

Here, we are studying both surfaces of one particular single slab. These results will be augmented in section 3 by studying the additional surfaces obtained by the variation of the slab thickness and density of defects initially contained in this latter.

Figure 3 presents the morphology, the inner strain cartography as well as the distribution of the sites according to the class they belong on the surface of the Top surface of the 5 layers slab with $ρ_v$=40%. In this very flat surface, in which some point vacancies are however noticeable, we realize that the surface is essentially made of two dimensional (2D) nano-domains with CN=7 (in red, that we shall denote in the followings as 7-domains) separated by linear chains of sites with CN=8 (blue, denoted as 8-chains), accommodating in between 2D 7-domains. The core of the 6-domains is composed of S-sites while D-sites (here in very negligible amount) are located at the vicinity of the point defects, these latter introducing a more disordered geometry. Sites with CN=6 and below are also observed close to these disturbed areas.

Figure 4 shows the second surface of the slab for which a more corrugated geometry were obtained. In such a situation, despite the 2D 7-domains are conserved, sites with CN=6 see their population increase to the detriment of the 7-domains (even if we cannot talk yet about 2D 6-domains for this particular slab). 8-chains are still observed and the role they play as accommodation pattern is confirmed: they separate 7-domains between each-others and 7-domains with 6-sites. Notice that S-sites essentially concern regions located in between (or far away from) the surface defects while D-sites constitute the periphery of the 7-domains, close to the defects, these latter gathering sites with more extreme and isolated CN (grey, violet and brown).

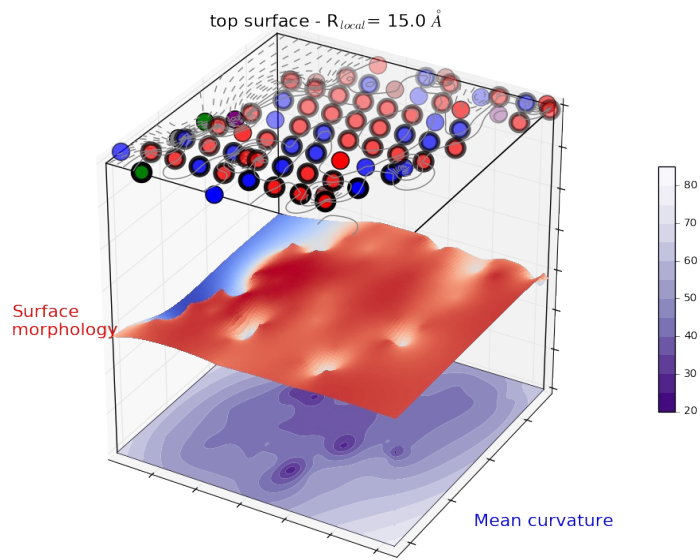

*Figure 3: In the first layer, sites of the surface are represented according to the category they belong (see text),. In the second layer, we display the morphology of the surface and in the last one, the cartography of inner strain. All these data correspond to the Top surface of the 5 layers slab with $\rho_v$=40%.*

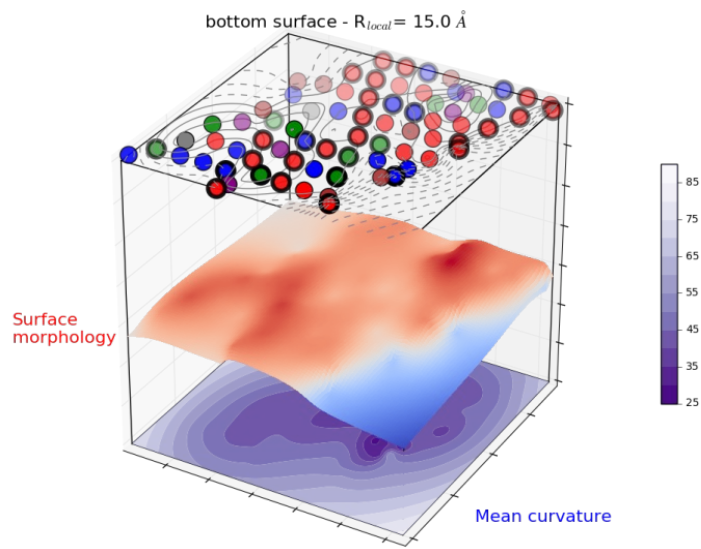

*Figure 4: same as Figure 3 but for Bottom surface.*

c. Electronic structure.

Figure 5 and Figure 6 report the projected density of states (DOS) of the surface sites of the top and bottom surfaces, respectively. If some noticeable discrepancies are observed between them due to the fine details of their local environment, two major classes of DOS are obtained for each CN. This point justifies *a posteriori* our sub-categorization of the classes of atoms in section by operating a mere splitting twice of the population above and below $\delta_c$=0.2 Å. For the Top surface, essentially made of 7-domains of S-sites, one type of DOS is recovered for CN=7.

For these latter, found in majority, the typical triangular shape of the DOS of the perfect Pt fcc-surface is identified. It is also the case for half the 6-sites (Figure 5-b): they correspond to S-sites. In Figure 5-a, the DOS of Disordered 6-sites are characterized by the development of a peak below the Fermi energy whose exact location depends on the exact local geometry of the distorted environment they undergo. Due to charge conservation, the appearance of such a peak close to the Fermi energy must be accompanied with an emptying of bands of lower energy, so that this type of sites rather bring a destabilization energy to the overall system. However they are found to be a natural option chosen by the system to adapt to the peculiar geometry imposed by the initial distribution of defects.

Regarding 8-chains, the more rectangular shape of the DOS of the D-sites (Figure 5-d) confirms the rather amorphous character of the 8-sites, identified, in the previous paragraph, as accommodation sites. The less numerous S-sites display more formed peaks due to the higher symmetry. However, the shape of the DOS is found in neat discrepancy with the one of perfect Pt surfaces, confirming the very particular status of these 8-sites, capable of absorbing the strain and accommodating between domains. Notice that it is made possible by the development of peaks at the bottom of the DOS of the S-sites, bringing stabilization energy to the system.

Figure 6 provides the DOS for the more defective bottom surface. Similar results are obtained. We nevertheless add them to evidence that a more defective surface presents the same patterns and fundamental characteristics: only the fraction of sites in each category varies.

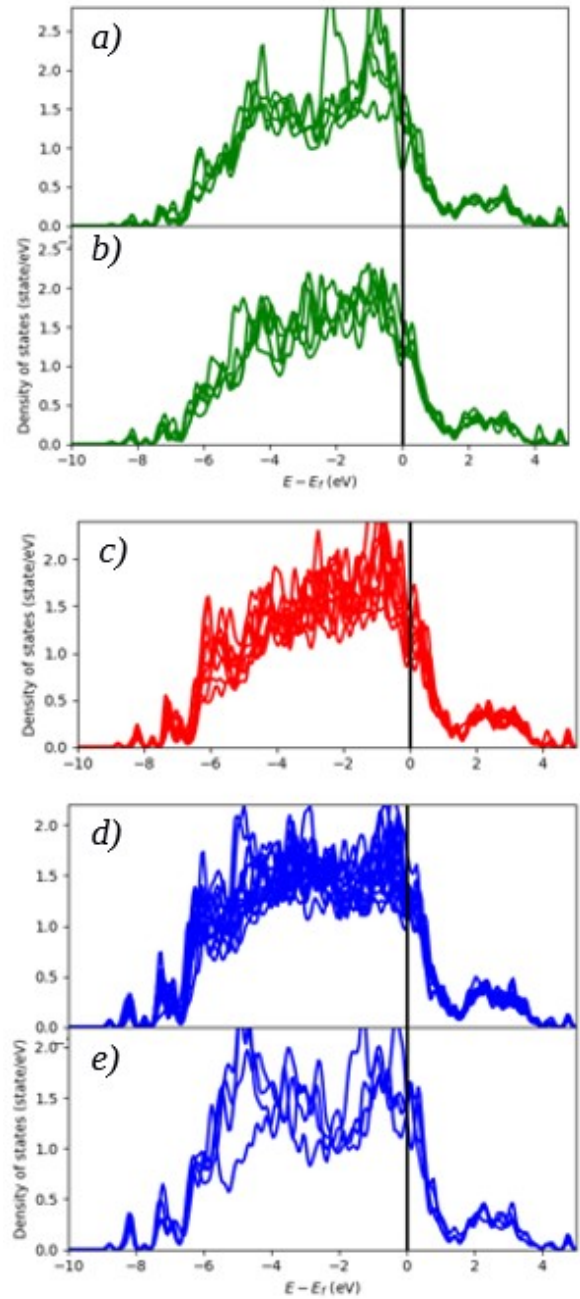

*Figure 5: Site-projected density of states of the surface sites belonging to the Top surface of the 5 layers slab with ρv=40%. According to our color code, green curves correspond to CN=6, red to CN=7 and blue ones to CN=8. Ef is the Fermi energy. It is considered as an absolute reference and fixed at 0 eV.*

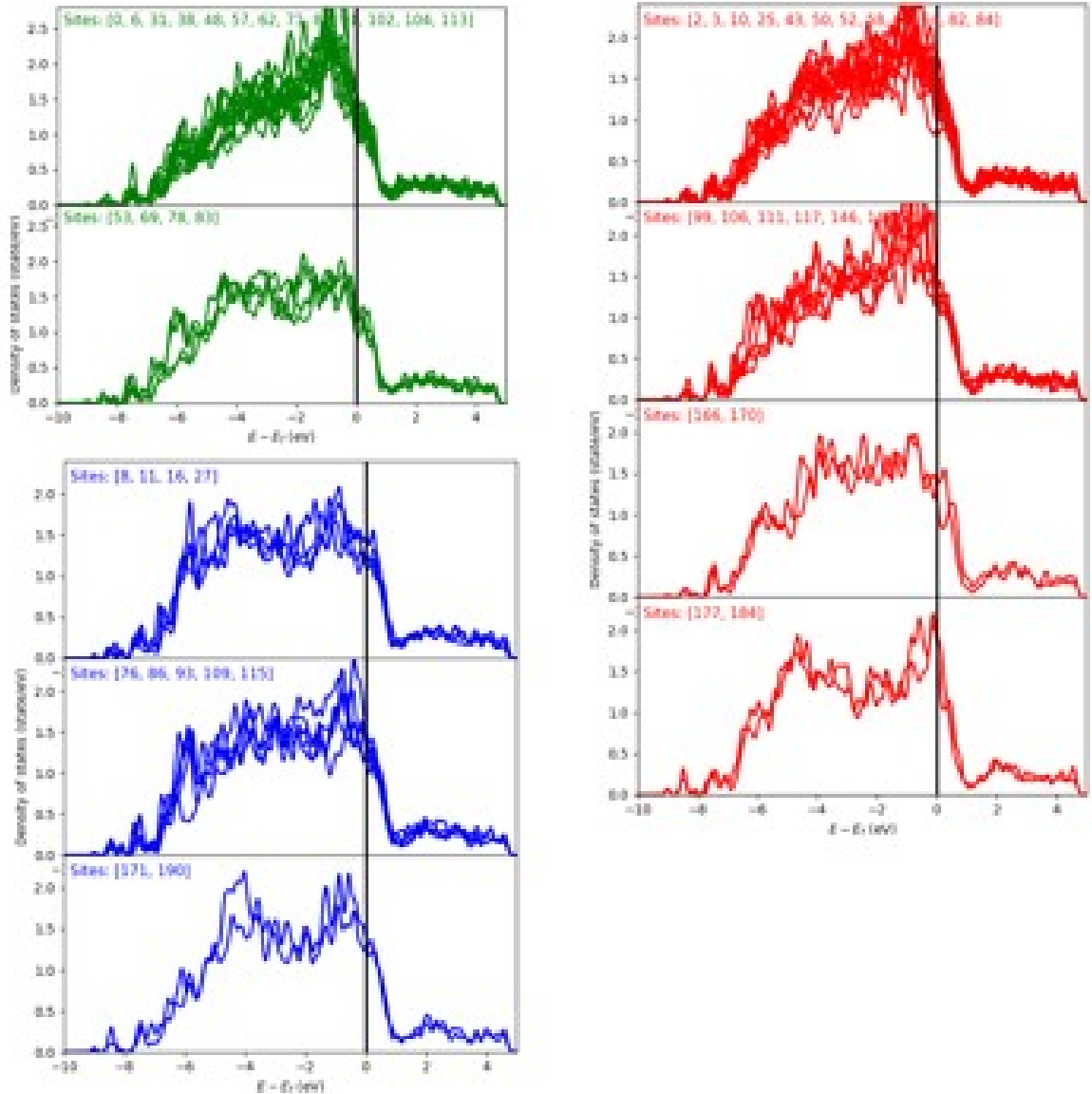

Figure 6: DOS of the Bottom surface.

## 3. Generalization to any morphology.

Let's now incorporate all the material acquired by the optimization of geometry of all our slabs involving 3, 5 and 7 atomic layers with various densities of defects (30%, 35%, 40%, 45% and 50%). The question we address here is how do these new surface morphologies affect the population in the various classes that we have determined for the 5 layers slab with $\rho_v$=40%.

Before investigating the evolution of the distribution of the domains as the thickness and amount of defects are varied, let's turn first, with Table I, to the change in the occurance per type of the sites in our new slabs.

| $\rho_v$ (%) | Surf. | 4 | 5 | 6 | 7 | 8 | 9 | 10 | Aspect | Plots |
|---|---|---|---|---|---|---|---|---|---|---|
| | | | | | CN | | | | | |
| [111] 5 layers (8x8 surface) | | | | | | | | | | |
| 30 | top | 0 | 1 | 7 | 45 | 36 | 6 | 1 | | |
| | bottom | 0 | 4 | 22 | 41 | 22 | 6 | 1 | | |
| 35 | top | 0 | 2 | 15 | 48 | 28 | 5 | 0 | | |
| | bottom | 2 | 3 | 18 | 45 | 22 | 7 | 0 | | Figure 7-c |
| 40 | top | 1 | 2 | 2 | 53 | 32 | 6 | 0 | Entirely flat | Figure 3 |
| | bottom | 2 | 6 | 13 | 42 | 24 | 9 | 1 | Flat + humps | Figure 4 |
| 45 | top | 0 | 6 | 24 | 39 | 21 | 7 | 0 | | Figure 7-a |
| | bottom | 0 | 4 | 13 | 40 | 29 | 11 | 0 | | |
| 50 | top | 2 | 5 | 17 | 48 | 24 | 2 | 0 | | |
| | bottom | 0 | 3 | 11 | 55 | 27 | 1 | 0 | | |
| [111] 3 layers (8x8 surface) | | | | | | | | | | |
| 40 | top | 8 | 10 | 23 | 47 | 8 | 1 | 0 | | |
| | bottom | 4 | 10 | 31 | 39 | 13 | 1 | 0 | | Figure 7-b |
| [111] 7 layers (8x8 surface) | | | | | | | | | | |
| 30 | top | 1 | 7 | 23 | 35 | 20 | 7 | 2 | | Figure 7-d |
| | bottom | 2 | 4 | 14 | 39 | 25 | 12 | 0 | | |

*Table I : The number of surface sites having a given CN is reported for the three thickness we have considered and the initial density of defects introduced in the slab. As all the slabs are made of a 8x8 [111] surface, the number of atoms in a surface is approximately the same for all the slabs. Approximately since the reconstruction of the surface combined with the density of defects may allow some initial atoms of surface to deep into a bulk layer.*

The first observation is that the number of sites with CN=7 remain majoritary whatever the slab may be. Their number vary slightly but quite probably as an effect of the initial implementation of the point defects made at random. Populations with CN=6 and 8 comes in more or less in the same proportion as our previous reference, the slight differences arising from the same reason as for CN=7. In general, it appears that the same phenomena of reconstruction are observed for our slabs, justifying a posteriori our focus on the 5 layers slab with $\rho_v$=40% to draw the main trends

More interestingly, Fig. 7 displays the geometrical arrangement of the sites per type for four selected slabs. The most striking point stands in the shrinking of the 2D red domains corresponding to CN=7, even if they subsist in small domains. This phenomena has to be related to shape or morphology of the surface. In Fig. 3, these domains match with flat regions which is not the case for our new cells due to the implementation at random of the defects. Thus, it appears that the size of the area of 2D domains strongly depends on the shape of the surface induced by the reconstruction, and that this latter tends to be shrunk as the surface bends. Sites with CN=8 (blue) conserve their linear arrangement and still accommodate between the regions of different nature. Regarding the sites with CN=6 (green) the already observed trend to create slim 2D domains intensifies. They grow proportionally to the shrinkage of the bent 7-domains. This is the option found by the surfaces to compensate the inability of the 2D 7-domains to persist when the surface is corrugated in a stronger manner.

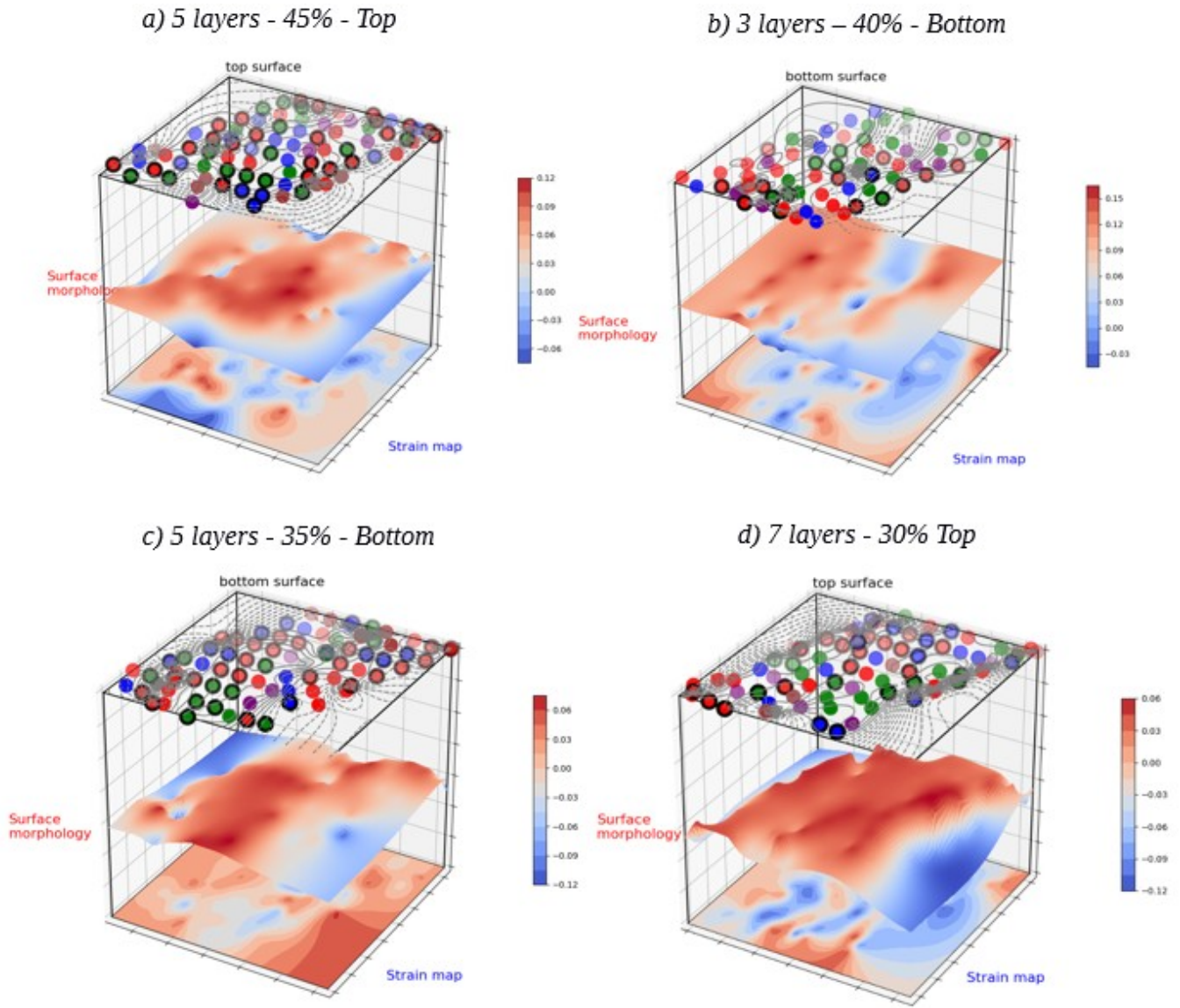

Figure 7: I same as Figure 3 but for a selection of new slabs among the ones we have investigated .

## 4. Adsorption energies of OH.

a. An apparent chaotic behavior, but a linear dependence revealed by the local geometry.

In their original paper [14], Calle-Vallejo et al. already demonstrated that a clear dependence of the adsorption energy of OH on flat facets of Pt belonging to small clusters as a function of the coordination number of the site couldn't be achieved. The introduction of an effective coordination number were found necessary to recover a neat and linear dependence of the adsorption energy of OH. The type of effective CN (noted cn) relying only on the influence of the number of atoms in the second shell of neighbors scaled by the maximum CN were found suitable in their study case, given the highly symmetric clusters they studied and consequently the negligible distortions of the sites.

In order to adapt such a procedure to our distorted surfaces, an implicit usage of the distribution of distances all around the central site has to be imagine in replacement of the cn. In our study case, these distributions are wide and their analysis shows that no frontier could be drawn between the first and second neighbors, rendering impossible the counting of the number of first and second neighbors. Figure 8.a shows, as in the case of clusters, the impossibility to extract a clear law for our systems as far as the coordination number is considered. Our only option is to award an appropriate weight for each atoms belonging to the sphere of neighbors. The radius of such a sphere has to be determined. This latter does not really necessitate to reach distances generally encountered for the second neighbors since the effects of these latter is indirectly included in the shape of the distribution of distances of the closest atoms. The radius has to be fixed so that any atoms actually influencing the energy of adsorption be selected, leaving the spectra of neighboring distances makes the job to fix the value of the weight. A value of *2.9 Å* were found the appropriate choice. A lower value were found ineffective, a larger value bringing no improvements.

The choice of the function defining the local geometry variable as a function of the interatomic distances were essentially driven by its upmost transferability or ability to recover a linear dependence of energy of adsorption as a function of this latter with a minimum of fitting parameters. Our choice turned to

$$cn_i = \alpha \sum_{j=0}^{n_j} e^{-\lambda \frac{d_{ij}}{a}} \qquad (2)$$

where $cn_i$ is our effective coordination variable centered on site i, alpha and lambda our fitting parameters, $d_{ij}$ the interatomic distances between the selected site, i, and its neighbors, j, $n_j$ the number of atoms included in the sphere of coordination and, a, the average lattice parameter in the surface plane. This latter scaling parameter for the interatomic distances were found mandatory to recover an acceptable linear behaviour for the energies as a function of $cn_i$. The result of this fit is displayed in Fig 8.b. The introduction of the a parameter as well as the combined usage of the exponential function appears to be crucial to reduce to allow a similar weight for the atoms whose distances are close below and above the average average inner lattice parameters, to bring a strong weight to the neighbors with very small interatomic distances and gradually attenuate the influence of the most distant neighbors.

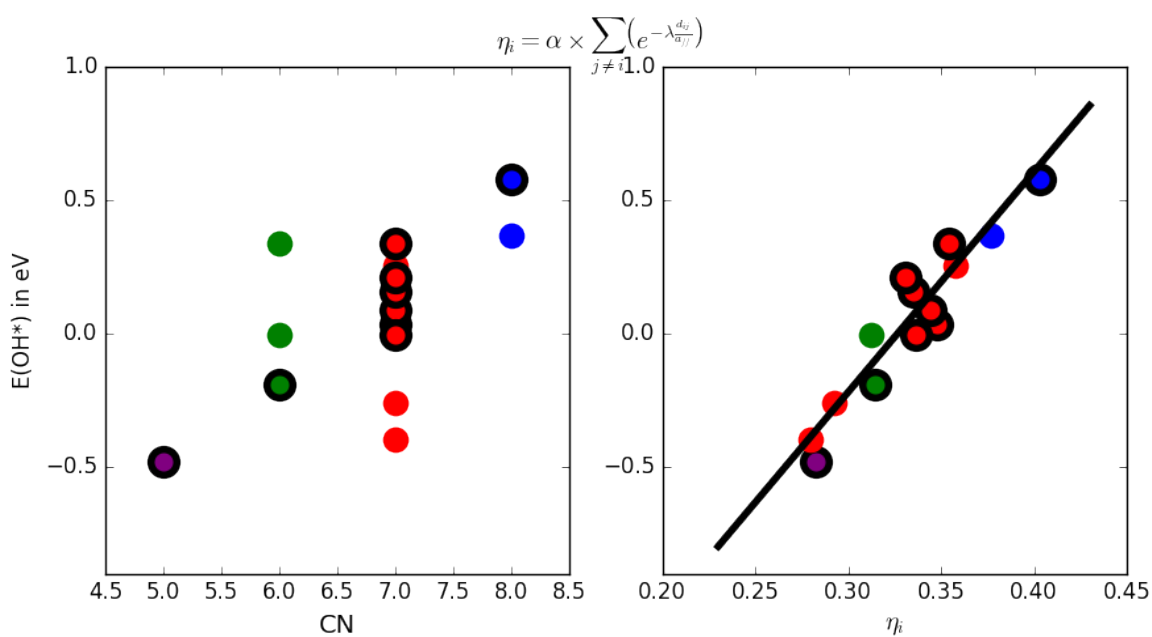

Figure 8: In a), we display the adsorption energies of OH on selected surface sites of both bottom and top surfaces of the 5 layers slab with $\rho_V$=40% as a function of the coordination number (CN) of the site. In b) they are plotted as a function of our effective parameter, $\eta_i$, taking into explicitly consideration the local environment of the site, namely the data of the interatomic distances between the neighboring atoms and the considered site within a cut-off of 2.9 Å (see the title of the present plot for the expression of $\eta_i$)

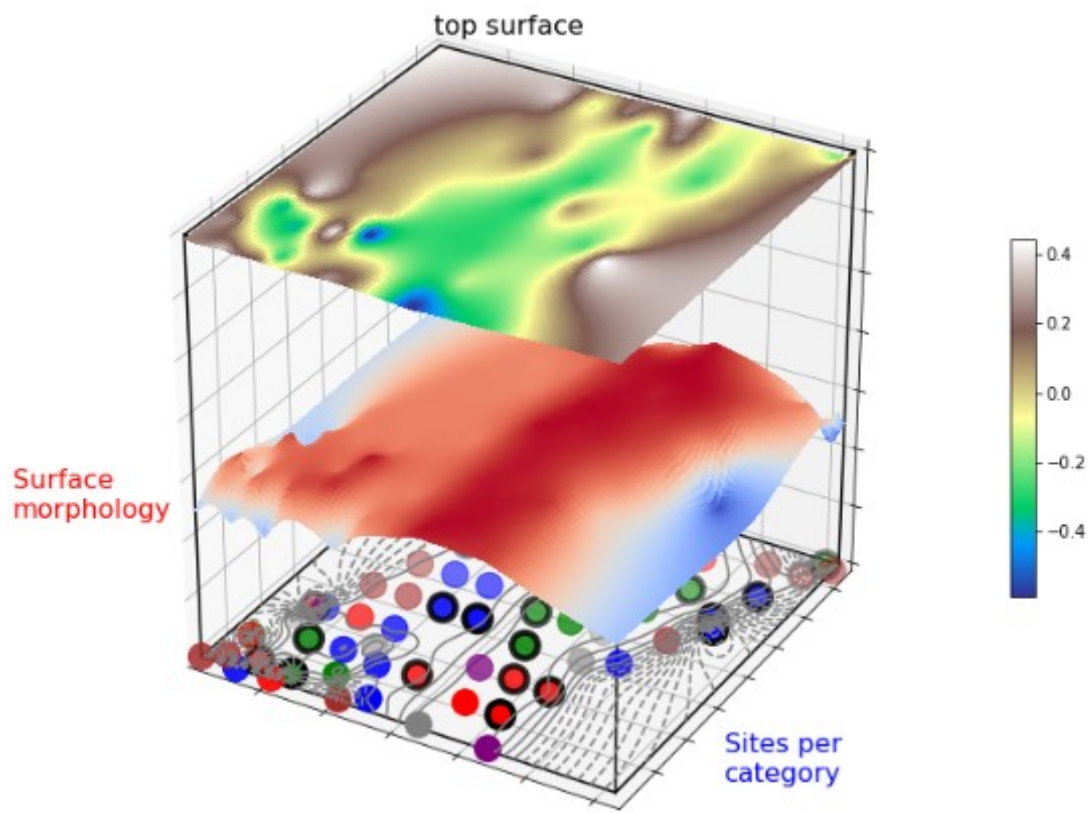

*Fig. 9 : in the top layer we present the absorption energy surface (in eV) calculated by means of the law of paragraph 4. This latter is only computed by the data of the interatomic distances obtained after relaxation of the slab. The morphology of the surface (middle layer) is added as well as the distribution per type of the various sites in the bottom layer.*

## Conclusion.

In this work, we made intensive use of DFT calculations to rationalize the catalytic activity of very distorted surfaces of Pt nanomaterials. Our investigations led us to realize that the various inequivalent sites of such surfaces may be classified into families according to their coordination number. Such a classification allowed us to determine that a specific role could be awarded to each family as far as the morphology of the surface is considered : 2D domains and rather flat regions concern the sites with a CN=7, while sites with CN=6 or 8 are found to arrange under the form of linear chain, accommodating in between the 2D domains. The other sites appears as isolated and sacrificed as far as we consider the crystallographic point of view. Let us recall that the size of the area of 2D domains strongly depends on the shape of the surface induced by the reconstruction, and that this latter tends to be shrunk as the surface bends. This step were found mandatory to select specific sites to study the reactivity of the surfaces : a reasonable number of sites had to be considered to provide a representative panel of behavior by means of a minimum of these very time-consuming calculations. From these data, a law piloting the values of the adsorption energies of OH as a function of the local geometry has been derived. Relying on two single fitting parameters and matching the DFT points with exactitude, a great confidence seems to be awarded to this latter to predict the energies of adsorption by the single knowledge of the distributions of the interatomic distances of the catalytic sites. We provide in Figure 9 a typical example based upon one of our slab. The major result is that the most active reactivity at the surface area (green region) gathers sites of various CN and, in that sens, does not superimpose with the cartography of the site types. Here, we argue that the fine step by step arrangement of the interatomic distances when the surfaces are reconstructed plays the key-role to determine the reactivity of the sites. Even if sites present a different CN, the fine local geometry of neighboring surface sites tends to accumulate their effects and create large regions of similar reactivity. This behavior was *a-priori* not expected.

## Acknowledgment.

We acknowledge GENCI under Project N° INP2227/72914 as well as PHYNUM CIMENT for computational resources. This work was performed within the framework of the Center of Excellence of Multifunctional Architectured Materials "CEMAM" n°ANR-10-LABX-44-01 funded by the "Investments for the Future" Program.

## Bibliography.